# Theoretical treatment of the interaction between two-level atoms and periodic waveguides

Xiaorun Zang, and Philippe Lalanne[*]

*Laboratoire Photonique Numérique et Nanosciences, Institut d'Optique d'Aquitaine, Université Bordeaux, CNRS, 33405 Talence, France.*
* E-mail: philippe.lalanne@institutoptique.fr

**Abstract**

Light transport in periodic waveguides coupled to a two-level atom is investigated. By using optical Bloch equations and a photonic modal formalism, we derive semi-analytical expressions for the scattering matrix of one atom trapped in a periodic waveguide. The derivation is general, as the expressions hold for any periodic photonic or plasmonic waveguides. It provides a basic building block to study collective effects arising from photon-mediated multi-atom interactions in periodic waveguides.

**Keywords**: Quantum description of interaction of light and matter, Atom traps and guides, Atoms in optical lattices, Optical implementations of quantum information processing and transfer, Quantum optics, Scattering, Nanophotonics and photonic crystals.

Introduction

Nanophotonic structures are routinely used to enhance light matter interactions by modifying the density of virtual photon states. Recently, new hybrid quantum systems, combining ultracold atoms and nanostructured devices, are emerged in order to attain new paradigms in quantum measurement and processing. The combination of excellent quantum coherence properties with a very flexible platform for implementing strong interactions at subwavelength scales is expected to go beyond classical settings of all-solid-state QED with quantum dots [1]. Examples include ultra-thin unclad optical fibers [2–4], photonic-crystal cavities formed from planar dielectrics [5], and more recently photonic periodic waveguides, for which the strong field enhancement at the band edge bears a high potential for atom-photon interaction [6–8].

The core physics behind all-optical quantum manipulation with periodic waveguides is the scattering of single photons by single atoms trapped near the waveguide, which is characterized by the scattering matrix of one atom [9]. Here we present a theoretical derivation of this matrix and propose semi-analytical expressions that can be easily calculated. The derivation accurately considers the phase of transmitted and reflected photons, the saturation of the atom for multiphoton incident states and radiation due to imperfect atom-waveguide coupling; it represents a first step towards a quantitative analysis of photon multiple scattering transport in periodic waveguides.

We maintain the discussion at a general level, potentially with waveguides composed of lossy and dispersive materials, the dielectric geometry being a specific case. The sole assumption is that the waveguide is made of reciprocal material. The scattering-matrix derivation can be performed with a semi-classical formalism, but we rely here on a fully-quantum treatment including a quantization of the atom and photon field. The approach is based on a combination of electromagnetic Bloch-mode-expansion techniques [10] with the optical Bloch equations for the atom density-matrix operator [11]. The former provides an accurate electromagnetic description of the system, including the local electric field and local density of electromagnetic states, and the latter allows us to describe the population evolution of the quantum system, assumed to be a two-level system. Both formalisms are already documented in the literature, and we do not present them again hereafter. Rather, we focus on how they are coupled to obtain the scattering matrix expression. The derivation is inspired from earlier theoretical works on the coupling of quantum emitters with *translation-invariant* dielectric [9,12,13] or metallic [14,15] waveguides, and coupled-resonator guides [16,17].

### Quantum treatment

Let us assume that the atom can be considered as a two-level system with ground and excited states ($|g\rangle$, $|e\rangle$) separated by frequency $\omega_A$ and that it is initially in the ground state and is driven by a coherent laser field at frequency $\omega_L$ launched into a Bloch mode of the waveguide, as sketched in Fig. 1.

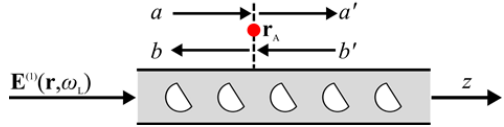

Fig. 1. Schematic of the hybrid system. An atom at $\mathbf{r}_A$ trapped near a periodic waveguide (made from reciprocal materials) and initially in its ground state. It is driven by a guided multiphoton coherent state, launched into the waveguide as Bloch mode $\mathbf{E}^{(1)}(\mathbf{r},\omega_L)$ propagating forward at frequency $\omega_L$. The scattering matrix for one atom [Eq. (6)] relates the scattered amplitudes, $a'$ and $b$, of the Bloch mode to the incident ones, $a$ and $b'$.

The interaction between the photons and the atom is described by a classical Hamiltonian

$$\hat{H} = \int d^3\mathbf{r}\int_0^\infty \hat{\mathbf{f}}^+(\mathbf{r},\omega)\hat{\mathbf{f}}(\mathbf{r},\omega)\hbar\omega d\omega \\ + \frac{1}{2}\hbar\omega_A\hat{\sigma}_z - \left[\boldsymbol{\mu}\cdot\hat{\mathbf{E}}^-(\mathbf{r}_A)\hat{\sigma} + \hat{\sigma}^+\hat{\mathbf{E}}^+(\mathbf{r}_A)\cdot\boldsymbol{\mu}\right] \tag{1}$$

The first term describes the electromagnetic field and the media, with $\hat{\mathbf{f}}(\mathbf{r},\omega)$ and $\hat{\mathbf{f}}^+(\mathbf{r},\omega)$ the bosonic vector field operators for the elementary excitations of the system; the second term accounts for the energy of the atom, with $\hat{\sigma}_z = |e\rangle\langle e| - |g\rangle\langle g|$ being the population difference operator; the third term accounts for the interaction between the atom and the total electric field, $\boldsymbol{\mu}$ being the transition electric dipole moment and $\hat{\sigma}$ the coherence operator. The electric field operator $\hat{\mathbf{E}}^+(\mathbf{r}_A)$ at the position $\mathbf{r}_A$ of the atom consists of three contributions: the vacuum field operator, the quantized emission field of the atom, whose main ingredients are driven by the imaginary part of the classical Green tensor of the periodic waveguide and the bosonic vector field operators, and the externally driving field. By assuming that the driving field is not just a single photon, but a coherent state at frequency $\omega_L$ that can treated as a classical complex field, the nonlinear response of the atom can be formally mapped to an external Rabi frequency into the optical Bloch equations, see details in [18] and in the method section in [14]. Under the rotating wave approximation, the dynamics of the expectation values is given by optical Bloch equations and the expectation value of the total (coherent) field operator at the laser frequency can be cast into two terms [11]

$$\langle\mathbf{E}(\mathbf{r})\rangle = \mathbf{E}^{(1)}(\mathbf{r}) + \langle\mathbf{E}_{at}(\mathbf{r})\rangle, \tag{2}$$

where $\mathbf{E}^{(1)}(\mathbf{r})$ is the classical incident Bloch-mode field and $\langle\mathbf{E}_{at}(\mathbf{r})\rangle$ is the expectation value of the field radiated by the atom. This general result that is valid for weak atom-field coupling regimes is not restricted to our specific waveguide geometry [11]. If one further neglects the frequency shift (the Lamb shift in a vacuum electromagnetic environment) due to the self-action of the induced dipole and decoherence effects leading to time-dependent effective dipole moment, the Rabi frequency depends only on the driving laser field at the atom position and the expectation value for the induced dipole moment reads as

$$\langle\mathbf{d}\rangle = \frac{-2|\boldsymbol{\mu}|^2}{\hbar}\mathbf{E}^{(1)}(\mathbf{r}_A,\omega_L)\frac{(2\delta - j\gamma)}{4\delta^2 + 2|\Omega|^2 + \gamma^2}, \tag{3}$$

where $\delta = \omega_L - \omega_A$ is the detuning between the laser and atom frequencies, $\gamma = 2\omega_L^2|\mu|^2/(\varepsilon_0\hbar)\mathbf{u}^t \cdot Im(\mathbf{G}(\mathbf{r}_A,\mathbf{r}_A,\omega_L)) \cdot \mathbf{u}$ denotes the modified spontaneous decay rate of the excited state [11], and $\Omega = (2\mu/\hbar)\mathbf{u} \cdot \mathbf{E}^{(1)*}(\mathbf{r}_A,\omega_L)$ is the complex external Rabi frequency. In the previous expressions, $\mathbf{u}$ denotes the polarization unit vector ($|\mathbf{E}^{(1)}|\mathbf{u} = \mathbf{E}^{(1)}$) of the photon at the atom position and $\mathbf{G}(\mathbf{r},\mathbf{r}',\omega_L)$ is the Green tensor in the presence of the waveguide, corresponding to the electric field response at $\mathbf{r}$ to a point dipole current source at $\mathbf{r}'$. A phenomenological decay rate may be introduced in Eq. (3) to deal with dephasing decays, if necessary.

### Electromagnetic treatment

To solve for Eqs. (2) and (3), we expand the field radiated by the atom in the complete set of optical Bloch modes, including a discrete set of truly guided modes and a continuum of radiation modes. The expectation value of the total field operator is written as

$$z > z_A, \quad \langle \mathbf{E}(\mathbf{r}) \rangle = \sum_{p=1,N} t_p \mathbf{E}^{(p)}(\mathbf{r}) + \text{continuum}, \quad (4a)$$

$$z < z_A, \quad \langle \mathbf{E}(\mathbf{r}) \rangle = \mathbf{E}^{(1)}(\mathbf{r}) + \sum_{p=1,N} r_p \mathbf{E}^{(-p)}(\mathbf{r}) + \text{continuum}, \quad (4b)$$

where the $r_p$ and $t_p$ are the reflection and transmission amplitude coefficients. To calculate them, one relies on a biorthogonal form that allows us to handle the orthogonality between bound and continuum states. Such forms can be mathematically derived directly from Maxwell's equations with a formalism based on the Lorentz reciprocity theorem. The formalism and its difficulties to handle leakage with continuum states is well documented in textbooks on optical waveguides [19]. It was generalized to periodic waveguides in [10], by introducing complex spatial coordinate transforms that map the open problem with its associated continuum of radiation states to an approximated closed problem with a finite number of discrete states, called quasi-normal Bloch modes (QNBMs). The QNBM formalism and its biorthogonal forms have already been successfully used for several studies on periodic waveguides, for instance to calculate the coupling of quantum emitters with photonic-crystal waveguides [20], or tiny out-of-plane-scattering loss at waveguide terminations [21] or localization lengths of periodic waveguides with tiny imperfections [22].

We conveniently normalize the QNBMs such that $\int_S [\mathbf{E}^{(-p)}(\mathbf{r}) \times \mathbf{H}^{(p)}(\mathbf{r}) - \mathbf{E}^{(p)}(\mathbf{r}) \times \mathbf{H}^{(-p)}(\mathbf{r})] \cdot d\mathbf{S} = 4\mathcal{P}^{(p)}$, with $S$ any cross-section plane of the waveguide and $\mathcal{P}^{(p)}$ any real number. For truly-guided Bloch modes operating below the cladding light line and for transparent materials with $Im(\varepsilon) = 0$, $\mathbf{E}^{(-p)} = -(\mathbf{E}^{(p)})^*$ and $\mathbf{H}^{(-p)} = -(\mathbf{H}^{(p)})^*$, and $\mathcal{P}^{(p)}$ represents the Bloch-mode power flow. With this normalization, the scattering coefficients are directly obtained by considering that the atoms act as electric-dipole sources whose radiation emission feeds the waveguide QNBMs, and we have $t_p = (j\omega_L/4\mathcal{P}^{(p)})\langle\mathbf{d}\rangle \cdot \mathbf{E}^{(-p)}(\mathbf{r}_A) + \delta(p)$ and $r_p = (j\omega_L/4\mathcal{P}^{(p)})\langle\mathbf{d}\rangle \cdot \mathbf{E}^{(p)}(\mathbf{r}_A)$ [10], where $\delta(p)$ ($\delta(p)=1$ if $p=1$ and 0 otherwise) accounts for the incident background illumination of the atom. Using $\langle\mathbf{d}\rangle$ from Eq. (3), we obtain

$$t_p = \sigma_0 \frac{-j\varepsilon_0 c}{4\mathcal{P}^{(p)}} \frac{\gamma_0(2\delta_L - j\gamma)}{4\delta_L^2 + 2|\Omega|^2 + \gamma^2} \mathbf{E}^{(1)}(\mathbf{r}_A) \cdot \mathbf{E}^{(-p)}(\mathbf{r}_A) + \delta(p), \quad (5a)$$

$$r_p = \sigma_0 \frac{-j\varepsilon_0 c}{4\mathcal{P}^{(p)}} \frac{\gamma_0(2\delta_L - j\gamma)}{4\delta_L^2 + 2|\Omega|^2 + \gamma^2} \mathbf{E}^{(1)}(\mathbf{r}_A) \cdot \mathbf{E}^{(p)}(\mathbf{r}_A), \quad (5b)$$

where $\sigma_0 = 6\pi c^2/\omega_A^2$ denotes the extinction cross section on resonance of an isolated two-level system and $\gamma_0 = \omega_A^3|\mu|^2/(3\pi\varepsilon_0\hbar c^3)$ is the natural decay rate of the atom in a vacuum ($2\pi\hbar$ is Planck's constant, $\varepsilon_0$ the permittivity of vacuum and $c$ the light speed). Equations (5a)-(5b), which constitute the main result of the present work, represent general expressions that are valid for any photonic or plasmonic waveguides and for any two-level systems. They can be extended to incorporate dephasing and non-radiative decays [11].

Hereafter for simplicity, we focus on the amplitude coefficients of the incident Bloch mode only. The scattering matrix for one atom, which relates the scattered mode amplitudes, $a'$ and $b$, to the incident ones, $a$ and $b'$ (see Fig. 1), has the following form

$$\begin{pmatrix} a' \\ b \end{pmatrix} = \begin{bmatrix} t & r' \\ r & t' \end{bmatrix} \begin{pmatrix} a \\ b' \end{pmatrix}$$
$$= \begin{bmatrix} 1+\eta \mathbf{E}^{(1)}(\mathbf{r}_A) \cdot \mathbf{E}^{(-1)}(\mathbf{r}_A) & \eta \mathbf{E}^{(1)}(\mathbf{r}_A) \cdot \mathbf{E}^{(1)}(\mathbf{r}_A) \\ \eta \mathbf{E}^{(-1)}(\mathbf{r}_A) \cdot \mathbf{E}^{(-1)}(\mathbf{r}_A) & 1+\eta \mathbf{E}^{(1)}(\mathbf{r}_A) \cdot \mathbf{E}^{(-1)}(\mathbf{r}_A) \end{bmatrix} \begin{pmatrix} a \\ b' \end{pmatrix},$$
(6)

with $\eta = \sigma_0 \dfrac{-j\varepsilon_0 c}{4\mathrm{P}^{(1)}} \dfrac{(2\delta_L/\gamma_0 - j\gamma/\gamma_0)}{4(\delta_L/\gamma_0)^2 + 2(\Omega/\gamma_0)^2 + (\gamma/\gamma_0)^2}$.

Equation (6) has several implications:

- $t$ and $t'$ are equal (reciprocity).

- In sharp contrast to translation-invariant waveguides for which $\mathbf{E}^{(1)}(\mathbf{r}_A) \cdot \mathbf{E}^{(1)}(\mathbf{r}_A) = \mathbf{E}^{(-1)}(\mathbf{r}_A) \cdot \mathbf{E}^{(-1)}(\mathbf{r}_A)$ because of translation symmetry, $r$ and $r'$ are different in general. Thus the scattering matrices of periodic and translation-invariant waveguides are fundamentally different, and the formulas directly inspired from classical waveguide results in recent works [6, 23] clearly overlook the phase difference between $r$ and $r'$.

- For non-absorbent dielectric waveguides, since $\mathbf{E}^{(-1)} = -(\mathbf{E}^{(1)})^*$, $(r'/\eta)^* = r/\eta$ and $|r|^2$ and $|r'|^2$ are equal. Perhaps counter-intuitively, this holds regardless the atom location in the unit cell of the waveguide.

- For metallic waveguides such as nanoparticle chains that offer deep subwavelength transverse confinements [24], $|r|^2$ and $|r'|^2$ are different.

- For small driving fields, saturation is negligible and the spectral width is given by $\gamma$. As one tunes the band edge near the atomic transition line, the total decay rate is mainly driven by the coupling to the waveguide and becomes proportional to the group index [6].

For the sake of illustration, we consider a periodic nanowire formed by a chain of Si$_3$N$_4$ boxes in air, see the inset in Fig. 2. Despite the existence of large air gaps, remarkably low propagation losses of 2.1 dB/cm have been recently reported for this geometry at telecom wavelengths [25]. All dimensions are given in the caption of Fig. 2. For TE–like horizontal polarization, the nanowire waveguides support a truly–guided Bloch mode and the conduction-band-edge energy coincides with the Cesium D$_2$ transition line at 852 nm.

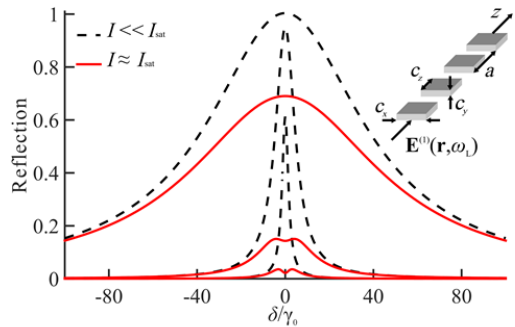

Fig. 2. Reflection spectra for an atom trapped in a sub-λ periodic nanowire (inset) for three values of the group index, $n_g$ = 3,10 and 80. Black dashed curves hold for low laser intensities, and red solid curves for a laser intensity close to the saturation intensity of Cs atoms. The calculation is performed for an atom in the middle of the air gap, and for $c_x$ = 500 nm, $c_y$ = 100 nm, $c_z$ = 220 nm and $a$ = 310 nm.

Figure 2 shows typical reflection spectra obtained for three values of the group index $n_g$ = 3, 10 and 80. At resonance and low excitation laser powers (black dashed curves), the photon is completely reflected for large $n_g$-values. As the band edge frequency is tuned near the atomic transition frequency, the atom serves as an ultranarrow filter and the incident laser field is almost completely reflected. The lineshape is the same as when a waveguide is coupled to a monomode cavity [26], but more complex Fano lineshapes can be obtained for multimode waveguides as can be deduced from Eqs. (5a)-(5b). The red solid curves are obtained for an averaged guided power of ≈ 5 pW. For this power that corresponds to 2.1 mW/cm², saturation effects cannot be neglected [27] and the coherent backscattering at the atomic transition frequency is reduced. At larger incident powers ($\Omega \gg \gamma_0$), most photons are directly transmitted, $|t|^2 \to 1$. Interestingly, we observe that saturation effects for the same averaged guided power are prominent at small $n_g$-values. This holds because the external Rabi frequency scales as $(n_g)^{1/2}$, whereas the spontaneous emission rate of the atom scales at a much faster rate, $\gamma \propto n_g$.

**Approximate treatment**

The main challenge for calculating the scattering matrix is due to the emission decay into radiation modes. This requires implementing accurate outgoing wave conditions in the periodic directions. Although PML-like absorbing boundaries in periodic media have been recently optimized [28], any termination breaks periodicity and numerical calculations relying on a 3D sampling is inevitably contaminated by termination backscattering, especially for small group velocities. Thus in practice, it is easier to consider finite waveguides. For instance, in a recent theoretical work [23], the classical Green's tensor is not calculated for fully-periodic (infinite) waveguides, but for finite ones formed by a finite number of unit cells. Consequently the predicted spontaneous decay rate displays a series of spurious resonance peaks, from which it is difficult to infer the actual decay rate of the periodic system.

We overcame the issue by using an approach that does not rely on numerical meshing in the periodic directions, but rather on an analytical expansion in the QNBM basis [10]. The approach provides the advantage to propose virtually "exact" predictions for $\gamma$, but in turn it relies on uncommon advanced numerical tools and on calculations that need to be repeated for every atom position. Thus it is advantageous to consider an approximate treatment, in which the spontaneous decay rate in all radiation Bloch modes $\gamma_{rad}$ is assumed to be equal to the emission rate in vacuum. Under this approximation,

$$\frac{\gamma}{\gamma_0} \approx 1 + \frac{3\pi c}{2\omega^2 \wp^{(1)}} \left|\mathbf{E}^{(1)}(\mathbf{r}_A)\right|^2 . \tag{7}$$

Together, Eqs. (5a)-(5b) and (6) allow us to analytically predict the spectral response for any driving frequency and atom location. The prediction only requires the knowledge of the driving electrical-field distribution, which is easily calculated for dielectric or metallic structures with Bloch-mode solvers.

Figure 3 provides an analysis of the error made with the approximate treatment as a function of the group index of the driving field for an atom trapped in the middle of the air gap and matched in frequency with the driving field. The approximate treatment slightly underestimates the actual backreflection. This is due to the fact that the decay into all radiation modes is smaller than the decay in vacuum ($\gamma_{rad} < \gamma_0$) [20], and we have checked that the "exact" data displayed in red solid in Fig. 3 can be perfectly reproduced by replacing 1 by 0.4 in Eq. (7). The approximate treatment offers a high degree of simplicity. For illustration, we calculate the reflection at zero detuning as a function of the atom position in the $x,z$-plane. The results are displayed in the inset of Fig. 3. They are likely to be inaccurate for small reflectance but the merit is to rapidly visualize the atom positions that matter for the coupling.

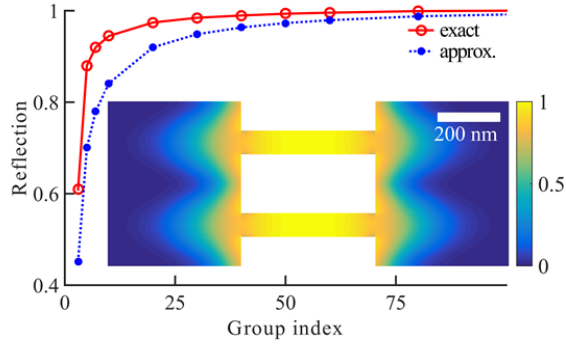

Fig. 3. Approximate treatment. Reflection spectra at low intensity for an atom in the middle of the air gap and matched in frequency with the driving field. The red solid and blue dotted curves are obtained with the "exact" and approximate treatments. The inset shows the reflectance as a function of the atom position in the ($x,z$) plane, predicted with Eq. (7) for $n_g = 40$.

## Conclusion

We have derived closed-form expressions for analysing the scattering of guided photons by cold atoms trapped near periodic waveguides. The expressions are generally valid for both photonic and plasmonic waveguides, the only restriction being that the waveguide material be reciprocal. The flexibility added by the analyticity is comfortable in practice, since it favors an in-depth understanding and allows a very efficient computation of the optical responses when some physical parameters are modified, such as the group velocity or the atom position. As a straightforward extension, by using Eq. (6) in conjunction with 2×2 Bloch-mode scattering-matrix-product algorithms [22], accurate and efficient calculations of light transport in periodic waveguides strongly coupled with a collection of dilute atoms may be achieved. This would help the interpretation of experiments aimed at measuring collective effects [8], which require ensemble-averaging over unknown atom positions. As a final remark, let us mention that Eq. (6) is not only valid for propagative incident photons in the band, but also for photons with energies within the gap (the normalization does not depends on energy consideration), implying that our theoretical treatment remains valid to study atomic interactions that are mediated by tunneling rather than propagating photons.

## Acknowledgments

The authors thank J. P. Hugonin for fruitful discussions and assistance. They also thank S. Bernon, P. Bouyer, B. Lounis and acknowledge financial support from Bordeaux Univ. and Région d'Aquitaine.